# Effect of annealing temperature on photovoltaic properties of spin coated CdS films

D. M. C. U. Dissanayake, J.S.T. Wickramasinghe and P. Samarasekara

Department of Physics, University of Peradeniya, Peradeniya, Sri Lanka

.

**Abstract**

CdS thin film samples were fabricated using spin coating technique on normal glass and conductive glass substrates and annealed at different temperatures. According to the XRD patterns of films prepared on normal glass substrates, polycrystalline films are consist of single phase of CdS. Films synthesized on conductive glass substrates were employed for photovoltaic measurements. Variation of photovoltaic properties with annealing temperature was investigated. The liquid junction photocell was measured in the electrolyte of $Na_2S_2O_3/I_2$. Then, the short circuit photo current and open circuit photo voltage were measured at room temperature under illumination and under dark using a solar cell simulation system. Optical band gap of CdS film samples was determined by measuring UV-visible spectroscopy.

## 1. Introduction:

CdS is a potential candidate in the applications of light emitting diodes, sensors, photoconductors, optical mass memories and solar selective coatings. CdS is a II-IV type semiconductor with a high band gap around 2.42 eV. Owing to the high band gap, CdS films are employed as a window material in CdS/CdTe solar cells. CdS is a n-type semiconductor with yellowish color. Nanostructures have attracted a great interest in recent years because of their unique chemical, physical, optical, electrical and transport properties. Owing to the high surface area, all nanostructured materials posses a high surface energy and thus, are thermodynamically unstable or metastable. CdS nanoparticles are considered to be a prime candidate in the applications in future opto-electronic devices, nanodevices and biological labeling due to availability of discrete energy levels, tunable band gap, size dependent chemical and physical properties, better chemical stability and easy preparation techniques.

Fabrication techniques of CdS thin films include spray pyrolysis [1], chemical bath deposition [2, 7,], electro deposition [3], screen printing [4], physical vapor deposition, vacuum evaporation [5], electron beam evaporation [6] and sol-gel spin coating. Among many deposition methods, sol-gel spin coating technique is extensively applied as a matrix material method to produce nanocomposites because it gives a higher specific surface area, superior homogeneity and purity, better microstructural control of metallic particles, narrow pore size and uniform particle distribution. The main advantages of the sol-gel method are its simplicity, low cost and its ability to obtain uniform films with good adherence and reproducibility in a relatively shorter processing time at lower sintering temperatures. Influence of sol aging time and annealing temperature on nanocrystalline CdS thin films has been investigated [8]. Previously, structural and optical characterization of sol-gel spin- coated nanocrystalline CdS thin film have been investigated [9].

Previously we have prepared film samples using chemical vapor deposition [10] and sputtering techniques [11, 12, 13] incorporated with expensive vacuum machines. However, spin coating technique was found to be low cost compared to above methods. CdS possesses some magnetic properties [14]. The Heisenberg Hamiltonian was used to describe the magnetic properties of ferromagnetic and ferrite films by us previously [15, 16, 17, 18, 19]. Previously band gap of semiconductor particles doped with salts have been investigated by measuring electrical properties [20]. In this manuscript, the variation of optical and structural properties of CdS films with number of layers will be described.

## 2. Experimental:

Initially two solutions have been prepared as following. Polyethylene Glycol (PEG) was dissolved in ethanol ($CH_3CH_2OH$), and acetic acid ($CH_3COOH$) was added to ethanoic solution under stirring which was continued for 1 hour. Cadmium nitrate ($Cd(NO_3)_2$) and thiourea ($CS[NH_2]_2$) were dissolved in ethanol under stirring which was continued for 1 hour. Two solutions were mixed and stirred again for 4 hours to obtain the final sols for deposition of thin films. CdS thin films were deposited on ultrasonically cleaned amorphous glass substrates by sol-gel spin-coating technique. Solution was applied onto the glass substrates at speeds of 1500, 2200 and 2400 rpm for 30 seconds. Thereafter, the samples were dried on a hot plate at 120 $^0$C for 1 hour, and annealed at 300, 350, 400, 450 and 500 $^0$C at five different annealing times of 20, 40, 60, 80 and 100 min in air. This method was repeated to fabricate multi layers of CdS films. CdS films with 10 layers were deposited.

Structural properties of film samples were determined using X ray diffraction (XRD) with Cu-K$_\alpha$ radiation of wavelength 1.54060 Å. UV-visible spectrometer Shimadzu UV1800 was employed to investigate the optical properties of samples. Photovoltaic properties were measured as follows. The photo cell was fabricated using CdS thin films spin-coated on fluorine-doped tin oxide (FTO) glass substrates as the working electrode, a solution of 1.0 mol dm$^{-3}$ of $Na_2S_2O_3$ as the liquid electrolyte and a Pt plate as the counter electrode. Then, the short circuit photo current and open circuit photo voltage were measured at room temperature under illumination and in dark using a solar cell simulation system (Portable Solar Simulator PEC-L01). The solar simulator illumination intensity on the electrode was 100 mWcm$^{-2}$ with voltage applied from -0.2 V to 0.2 V in steps of 0.01 V and cell active area was 1 cm$^2$.

## 3. Results and discussion:

Figure 1 shows the XRD patterns of CdS film samples with 10 layers annealed at 350, 400, 450, 500 $^0$C in air for 1 hour. All the samples given here were spin coated at spin speed of 2200 rpm for 30 s. Diffraction analysis suggests that all the films are polycrystalline in nature and crystallized in hexagonal wurtzite structure. These patterns exhibit prominent peaks corresponding to reflections (100), (002), (101), (110), (103), (112) together with low intensity peaks corresponding to reflections (102) and (201), and are in agreement with standard X-ray diffraction data reported in ICDD-PDF file 772306.

However, it is difficult to resolve the diffraction peak at 26.5$^0$ since cubic (111) and hexagonal (002) reflections coincided at this point. CdS thin films change from the metastable cubic zincblende phase to the stable hexagonal wurtzite phase by thermal annealing in the range 240-300 $^0$C. In this study, thin films were annealed starting from 300$^0$C and onwards. Therefore, can conclude that no phase transition occurred during annealing for any film, and the only existing crystal phase is hexagonal wurtzite phase. Further, films annealed at 350 $^0$C don't exhibit the latter two low intense peaks. Diffraction peaks become more shaper and their intensities found to be increased with annealing temperature. This may be due to improved crystallinity of films and more ordering of the films. To quantitatively investigate the preferred orientation, the degree of preferred orientation (σ) and the texture coefficient ($C_{hkl}$) were calculated for prominent peaks and are presented below in table 1and figure 2 as a function of the annealing temperature.

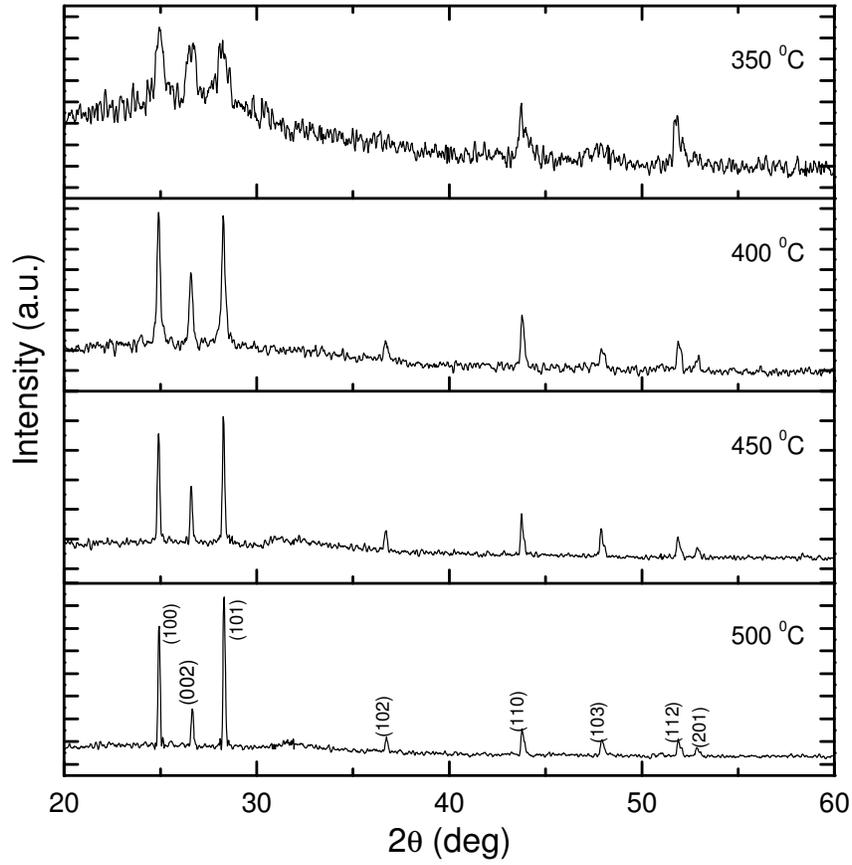

Figure 1: XRD patterns of CdS film samples with 10 layers annealed at 350, 400, 450, 500 $^{0}$C in air for 1 hour.

| Texture Coefficient | $C_{100}$ | $C_{002}$ | $C_{101}$ | $C_{110}$ | $C_{103}$ | $C_{112}$ |
|---|---|---|---|---|---|---|
| 350 $^{0}$C | 0.05 | 2.86 | 0.03 | 0.01 | 3.04 | 0.01 |
| 400 $^{0}$C | 1.33 | 1.17 | 0.94 | 0.98 | 0.51 | 1.08 |
| 450 $^{0}$C | 1.45 | 0.99 | 1.19 | 0.86 | 0.63 | 0.89 |
| 500 $^{0}$C | 1.42 | 0.79 | 1.50 | 0.85 | 0.46 | 0.83 |

Table 1: Texture coefficients of CdS thin films annealed at different temperatures.

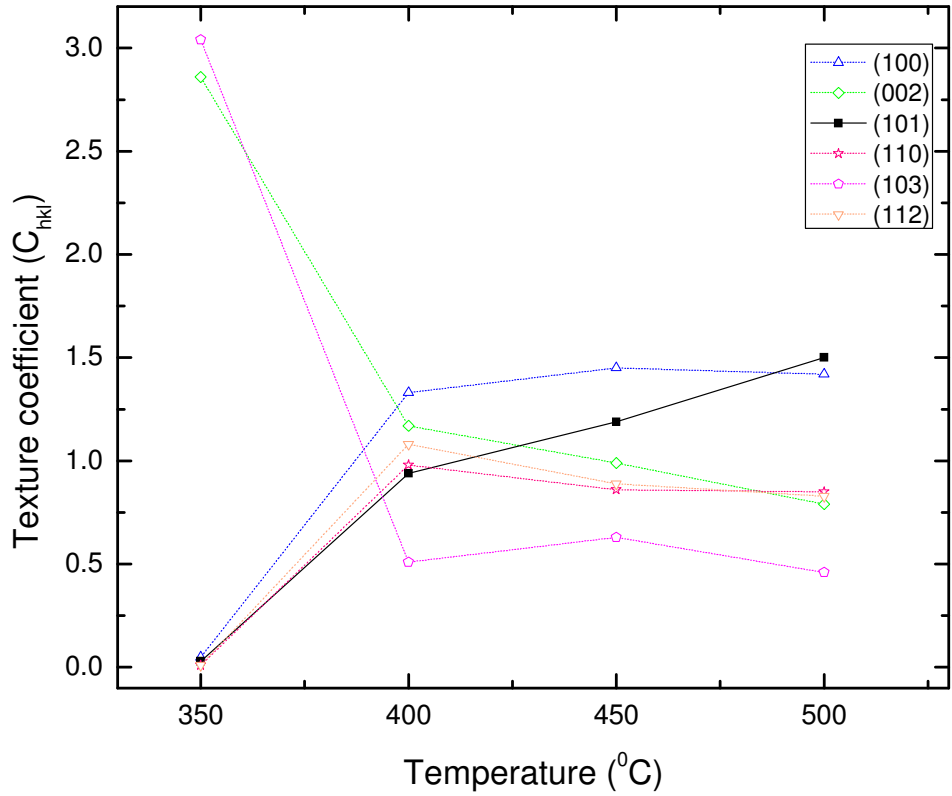

Figure 2: Variation of $C_{hkl}$ of CdS thin films with annealing temperature

The texture coefficient of a certain crystal plane (hkl) in a polycrystalline thin film is given by the relation,

$$C_{hkl} = \frac{I(hkl)/I_0(hkl)}{(1/N)[\sum_N I(hkl)/I_0(hkl)]} \quad (1)$$

Where,　　　h,k,l  = Miller indices corresponding to the diffraction peak

　　I(hkl)  = Measured intensity

　　$I_0$(hkl) = JCPDS standard intensity of the corresponding powder

　　N　　 = Number of reflections

The results point out that the texture coefficient of (101) plane increases with increasing temperature while other planes exhibit an oscillation in its value with annealing temperature. Therefore, it should be noted that the (101) plane of the hexagonal CdS structure is the most thermodynamically favorable growth plane, because it offers the lowest surface energy and hence the highest texture coefficient value.

Table 2 shows the variation of the degree of preferred orientation ($\sigma$) of CdS thin films with respect to annealing temperature.

| Temperature($^0$C) | $\sigma$ |
|---|---|
| 350 | 1.38 |
| 400 | 0.36 |
| 450 | 0.36 |
| 500 | 0.55 |

Table 2: Variation of the degree of preferred orientation ($\sigma$) of CdS thin films with annealing temperature.

The value of $\sigma$ can be calculated as following.

$$\sigma = \sqrt{\frac{\sum_{i=1}^{N}(C_{hkl} - C_{0hkl})^2}{N}}$$

Where, $\sigma$ = Degree of preferred orientation

$C_{0hkl}$ = Texture coefficient of powder sample which is always unity

UV absorption spectrums and Tauc plots were used to calculate optical band gap. Band gap values calculated for different annealing temperatures are given in Table 3, and these values are slightly higher than the theoretical band gap of CdS (2.42 eV). This may due to the formation of nano particles. When particle sizes become smaller band gap increases because of the quantum size effect. Therefore, the values obtained for optical band gap are higher than that of the band gap of bulk CdS.

| Temperature ($^0C$) | Bandgap (eV) |
|---|---|
| 300 | 2.494 |
| 350 | 2.461 |
| 400 | 2.431 |
| 450 | 2.434 |
| 500 | 2.434 |

Table 3: Optical band gaps at different annealing temperatures.

Figure 3 represents I-V characteristics of CdS thin films annealed at different temperatures. The linear nature of I-V curves indicates that the conduction mechanism in these films is ohmic.

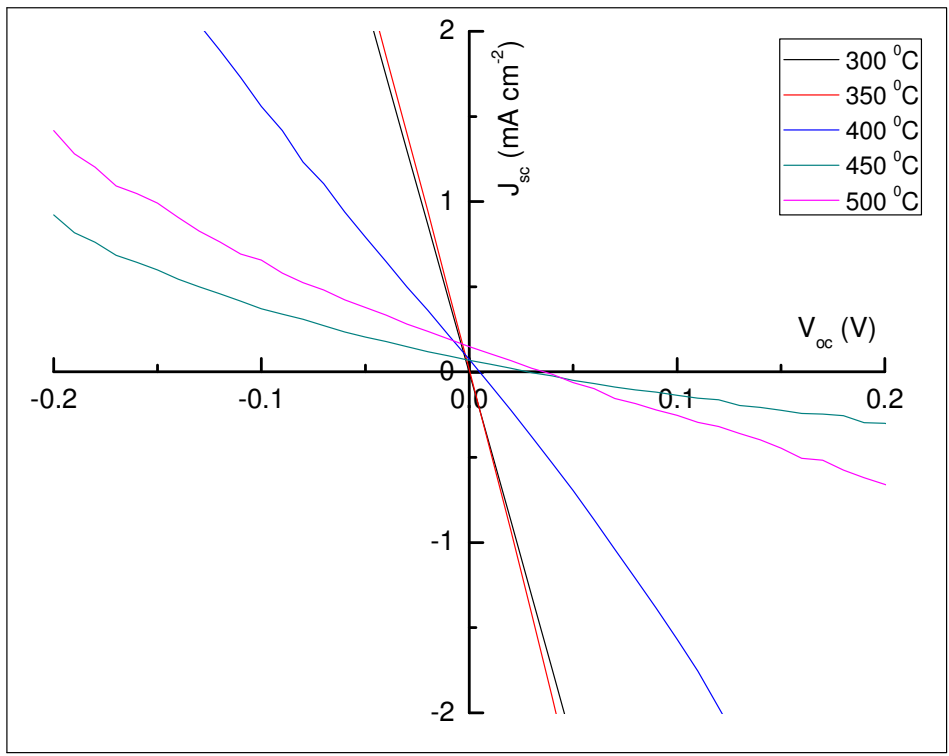

Figure 3: Current-Voltage characteristic of CdS thin films annealed at different temperatures

Following Table 4 summarizes photo current density, dark current density, photo voltage and photosensitivity.

| Temperature ($^0$C) | Photo current (μA cm$^{-2}$) | Dark current (μA cm$^{-2}$) | Photovoltage (X 10$^{-1}$ mV) | Photo sensitivity |
|---|---|---|---|---|
| 300 | 6.12 | 1.30 | 1.40 | 0.79 |
| 350 | 14.45 | 1.50 | 3.08 | 0.90 |
| 400 | 73.00 | 1.95 | 49.98 | 0.95 |
| 450 | 69.75 | 3.13 | 280.33 | 0.96 |
| 500 | 149.44 | 4.20 | 371.64 | 0.97 |

Table 4: Photo current, dark current, photo voltage and photo sensitivity of CdS thin films with annealing temperature.

Although both photo current and dark current increase with the annealing temperature, photo current is really high compared to the dark current. This phenomenon exhibits the positive photo conductivity of the material and because of the increase of generation of mobile charge carriers and their life time upon radiation. Another observation can be made from above Table 4 is that photosensitivity increases with annealing temperature due to the increase of charge carrier density and their mobility upon annealing. This result makes CdS thin films a capable candidate in photosensor devices. The variation of photo current density and photo voltage as a function of annealing temperature is shown in Figure 4.

According to figure 4, the short circuit photo current density increases with annealing temperature. This may be due to the fact that optical band gap decreases with annealing temperature allowing large amount of incident radiation to be absorbed and consequently increases the photo current. However, there is a slight reduction of photo current at 450 $^0$C and increases at 500 $^0$C. This again reflects variation of optical band gap with increase of annealing temperature. Since there was a band gap increase at 450 $^0$C and due to the slight band gap change at 500 $^0$C, photo current fluctuates accordingly.

Similarly, open circuit photo voltage increases with increasing annealing temperature. This may be due to decrease of defects and improvement of crystallinity with annealing temperature and hence reduction of optical band gap.

The conversion efficiency and fill factor were found to be very low for films annealed at lower temperatures. Efficiencies of $4.48 \times 10^{-4}$ and $13.16 \times 10^{-4}$ were calculated for films annealed at 450 and 500 $^0$C, respectively. Further, fill factor values were found to be 0.23 and 0.24 for films annealed at 450 and 500 $^0$C, respectively. The increase of efficiency and fill factor values are due to the improvement of crystallinity of thin films with annealing temperature.

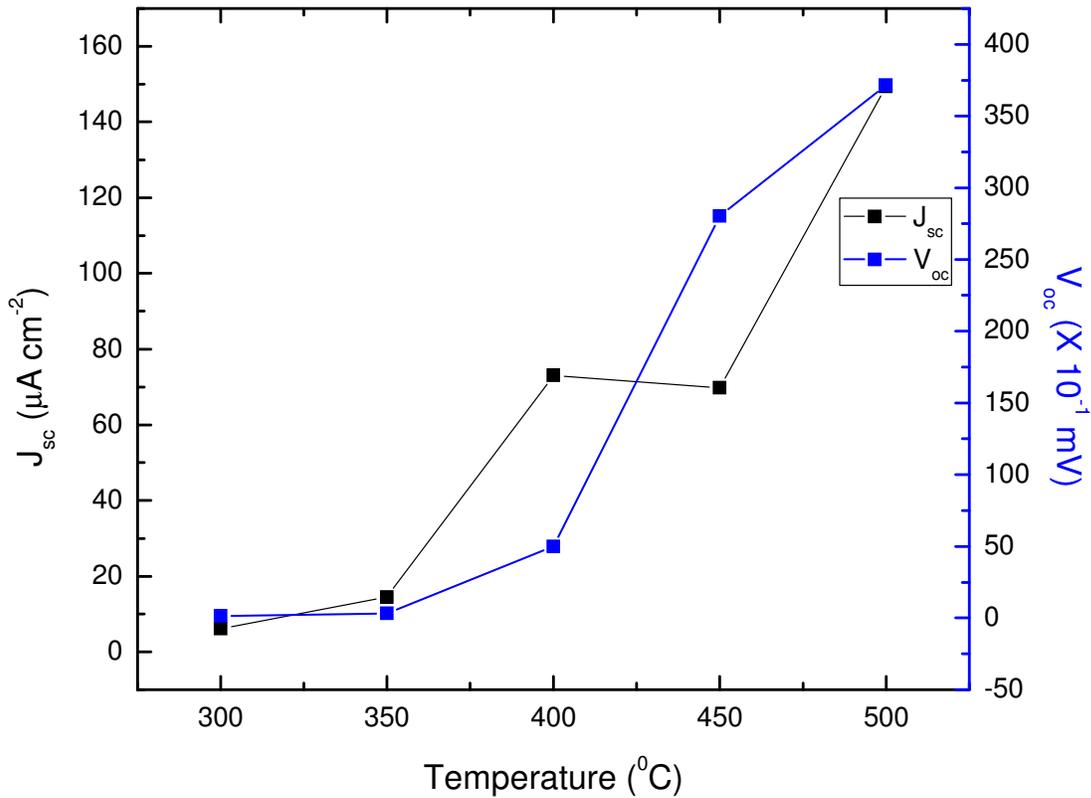

Figure 4: Variation of short circuit photo current density ($J_{sc}$) and open circuit photo voltage ($V_{oc}$) as a function of annealing temperature

## 4. Conclusions:

According to XRD patterns, crystallization of CdS films improves, as annealing temperature is increased from 350 to 500 $^0$C. Texture coefficient of (101) plane gradually increases from 0.03 to 1.50 with annealing temperature while other planes exhibit an oscillation in its value with increase of annealing temperature by implying that (101) is the thermodynamically favorable lattice plane. Optical bang gap calculated from UV-Visible spectroscopy and Tauc plots gradually decreases from 2.494 to 2.434 eV, as the annealing temperature is increased from 300 to 500 $^0$C. These band gap values are slightly higher than the theoretical band gap of CdS (2.42 eV) due to the formation of nano particles. All the photocurrent density, dark current density, photo voltage and photosensitivity increase with the annealing temperature. Photocurrent density increases from 6.12 to 149.44 µA/cm$^2$ as the annealing temperature is increased from 300 to 500 $^0$C. In addition, photosensitivity increases from 0.79 to 0.97, as the annealing temperature is increased from 300 to 500 $^0$C.

**References:**

1. A.A. Yadav, M.A. Barote and E.U. Masumdar, 2010. Studies on nanocrystalline cadmium sulphide (CdS) thin films deposited by spray pyrolysis. *Solid sate sciences* **12(7)**, 1173-1177.

2. P. Lisco, P.M. Kaminski, A. Abbas, K. Bass, J.W. Bowers, G. Claudio, M. Losurdo and J.M. Walls, 2015. The structural properties of CdS deposited by chemical bath deposition and pulsed direct current magnetron sputtering. *Thin solid films* **582**, 323-327.

3. M. Takahashi, S. Hasegawa, M. Watanabe, T. Miyuki, S. Ikeda and K. Iida, 2002. Preparation of CdS thin films by electrodeposition: effect of colloidal sulfur particle stability on film composition. *Journal of applied electrochemistry* **32**, 359-367.

4. V. Kumar, D.K. Sharma, M.K. Bansal, D.K. Dwivedi and T.P. Sharma, 2011. Synthesis and characterization of screen printed CdS films. *Science of sintering* **43(3)**, 335-341.

5. 1. K. Yilmaz, 2014. Some structural, electrical and optical properties of vacuum evaporated CdS thin films. *Journal of Ovonic Research* **10(6)**, 211-219.

6. Yang Dingyu, Zhu Xinghua, Wei Zhaorong, Yang Weiqing, Li Lezhong, Yang Jun, and Gao

Xiuying, 2011. Structural and optical properties of polycrystalline CdS thin films deposited by electron beam evaporation. *Journal of Semiconductors* **32(2)**, 023001 1-4.

7. Be Xuan Hop, Ha Van Trinh, Khuc Quang Dat, Phung Quoc Bao, 2008. Growth of CdS thin films by chemical bath deposition technique. *VNU journal of science, mathematics –Physics* **24**, 119-123.

8. I. Rathinamala, J. Pandiarajan, N. Jeyakumaran and N. Prithivikumaran, 2014. Synthesis and physical properties of nanocrystalline CdS thin films- Influence of sol aging time and annealing temperature. *International Journal of Thin Films science and Technology* **3 (3)**, 113-120.

9. M.A. Olopade, A.M. Awobode, O.E. Awe and T.I. Imalerio, 2013. Structural and optical characterization of sol-gel spin- coated nanocrystalline CdS thin film. *International journal of research and reviews in applied sciences* **15(1)**, 120-124.

10. P. Samarasekara, 2009. Hydrogen and Methane Gas Sensors Synthesis of Multi-Walled Carbon Nanotubes. *Chinese Journal of Physics* **47(3)**, 361-369.

11. P. Samarasekara, 2010. Characterization of Low Cost p-$Cu_2O$/n-CuO Junction. *Georgian Electronic Scientific Journals: Physics* **2(4)**, 3-8.

12. P. Samarasekara and N.U.S. Yapa, 2007. Effect of sputtering conditions on the gas sensitivity of Copper Oxide thin films. *Sri Lankan Journal of Physics* **8**, 21-27.

13. P. Samarasekara, A.G.K. Nisantha and A.S. Disanayake, 2002. High Photo-Voltage Zinc Oxide Thin Films Deposited by DC Sputtering. *Chinese Journal of Physics* **40(2)**, 196-199.

14. X.G. Zhao, J.H. Chu and Z. Tang, 2015. Magnetic properties, Heisenberg exchange interaction, and curie temperature of CdS nanoclusters. *The journal of physical chemistry* **119(52)**, 29075-29086.

15. P. Samarasekara and Udara Saparamadu, 2012. Investigation of Spin Reorientation

in Nickel Ferrite Films. *Georgian electronic scientific journals: Physics* **1(7),** 15-20.

16. P. Samarasekara and N.H.P.M. Gunawardhane, 2011. Explanation of easy axis orientation of ferromagnetic films using Heisenberg Hamiltonian. *Georgian electronic scientific journals: Physics* **2(6),** 62-69.

17. P. Samarasekara, 2008. Influence of third order perturbation on Heisenberg Hamiltonian of thick ferromagnetic films. *Electronic Journal of Theoretical Physics* **5(17),** 227-236.

18. P. Samarasekara and Udara Saparamadu, 2013. In plane oriented Strontium ferrite thin films described by spin reorientation. *Research & Reviews: Journal of Physics-STM journals* **2(2),** 12-16.

19. P. Samarasekara and Udara Saparamadu, 2013. Easy axis orientation of Barium hexa-ferrite films as explained by spin reorientation. *Georgian electronic scientific journals: Physics* **1(9)**, 10-15.

20. K. Tennakone, S.W.M.S. Wickramanayake, P. Samarasekara and, C.A.N. Fernando, 1987. Doping of Semiconductor Particles with Salts. *Physica Status Solidi (a)* **104**, K57-K60.